\newcommand{\Load}{{\bf F^{\rm ext}}}       
\newcommand{\Poly}{R_{\rm max}}              
\newcommand{\Resp}{{\bf G}(\alpha\beta|\gamma)} 

\documentclass[epj,referee]{svjour}
\usepackage{psfig}

\begin{document}

\title{Robust propagation direction of stresses in a minimal
granular packing}

\author{D. A. Head,\inst{1} A. V. Tkachenko\inst{2} and T. A. Witten\inst{3}
}

\institute{Department of Physics and Astronomy, JCMB King's Buildings,
University of Edinburgh, Edinburgh EH9 3JZ, UK
\and
Bell Labs, Lucent Technologies, 600-700 Mountain Ave.,
Murray Hill, NJ 07974, USA
\and
The James Franck Institute, The University of Chicago,
Chicago, Illinois 60637, USA}

\date{\today}

\abstract{
By employing the adaptive network simulation method,
we demonstrate that the ensemble-averaged
stress caused by a local force for packings of frictionless rigid beads
is concentrated along rays whose slope is consistent with unity:
forces propagate along lines at 45 degrees to the horizontal or vertical.
This slope is shown to be independent
of polydispersity or the degree to which the system is sheared.
Further confirmation of this result comes from fitting the
components of the stress tensor to the null stress `constitutive equation.'
The magnitude of the response is also shown
to fall off with the -1/2 power of distance.
We argue that our findings are a natural consequence of a
system that preserves its volume under small perturbations.
}
%
\titlerunning{Robust propagation direction of granular stresses}
\maketitle

\section{Introduction}
\label{s:intro}
It is becoming increasingly evident that the transmission of stress
through a disordered body of incompressible, cohesionless beads
obeys principles that are distinct from those of
either elastic or elastoplastic
materials~\cite{royalsoc,devils,exp_history,dip,exp_column,footprints}.
Even defining a strain field for a granular body
is problematic~\cite{royalsoc,devils}, let alone relating
the strain to the stress in a constitutive equation,
as in elastic systems.
Furthermore, experiments measuring the stress under conical and wedge--shaped
sandpiles have indicated a pressure minimum below the apex when the 
material is poured from a point (or line) source, but not when it is
sprinkled uniformly over the base~\cite{exp_history}.
This dependence on construction history is also quite unlike
elastic or elastoplastic materials.

To circumvent the difficulties arising from an ill--defined strain,
a class of `constitutive equations' have been proposed
that relate the independent components of the stress tensor $\sigma_{ij}$
without reference to a displacement field~\cite{royalsoc,OSL,FPA}.
(Note that, like many authors, we use the phrase `constitutive
equation' even in the absence of a well-defined strain field
or history independence).
In their most general form, these equations suppose that
the $\sigma_{ij}$ can be related in a linear expression,
with coefficients that in general depend on the
construction history of the pile.
The basic form of these equations
has also been confirmed by mean field analysis of
frictionless spheres, where it was referred to as the
`null stress' law~\cite{first_pre}.

A remarkable feature of the proposed equations is that
they are hyperbolic in form, and thus belong to the same class of
equations as the wave equation.
Hence they predict that the response to a localised perturbation
should propagate outwards from the source of the disturbance in waves,
with the direction of increasing depth playing the role of time.
To test this prediction, experiments have recently been
performed on systems of photoelastic discs that have a small
load applied to the upper surface, and the stress response
in the bulk measured~\cite{footprints}.
These seemed to show wave--like propagation over all distances
for low polydispersity, but only for short distances for high
polydispersity.
To further complicate the issue, similar experiments on sand
have failed to show any trace of wave--like propagation
whatsoever~\cite{greens_sand}.

Until more experiments have been performed and a consensus reached,
it is natural to turn to computer simulations for additional data.
However, conventional simulation methods cannot directly probe the
linear response regime and must instead apply a finite force,
which runs the risk of perturbing the system.
To this end, two of us (AT, TW) have devised a novel simulation method,
called the `adaptive network algorithm,' which directly probes the
linear regime in a system of frictionless and rigid discs~\cite{adapt_net}.

In this paper, we employ the adaptive network algorithm to
investigate the stress response function in frictionless disc systems,
and to ascertain if it obeys wave--like propagation.
We begin by briefly describing the simulation method in Sec.~\ref{s:model},
before demonstrating in Sec.~\ref{s:nullstress} that the null--stress law
is obeyed, with the stress tensor approximately obeying the relation
$\sigma_{yy}=c^{2}\sigma_{xx}$ with $c\approx1$.
This result has already been found~\cite{adapt_net}
for weakly polydisperse systems,
with the same value of~$c$; here we extend the result to strong
polydispersity.

The response function itself is analysed in Sec.~\ref{s:response},
where it is shown to be concentrated along two `light rays' that
propagate outwards from the source of the disturbance.
Furthermore, the same `speed of light' $c\approx1$ is found
in all cases, irrespective of the direction of the perturbing
force, the polydispersity of the system, or the degree to which
the system is being sheared.
This behaviour persists to the largest depths we have been
able to simulate, typically around 10---15 bead diameters
in the vertical direction,
which is comparable with the experiments on photoelastic discs;
however, they found no evidence of $c\approx1$.
In Sec.~\ref{s:disc} we attempt to explain this discrepancy
by arguing that $c=1$ corresponds to a system that conserves
its volume under small perturbations.
Finally, in Sec.~\ref{s:conc} we summarise our results
and suggest directions for future work.

\section{Overview of the simulation method}
\label{s:model}

The full derivation of the adaptive network algorithm and its associated
assumptions has already been presented at length elsewhere~\cite{adapt_net};
here we only mention those details pertinent to the current work.
The algorithm consists of two stages:
firstly, a packing of discs is constructed,
and secondly, the contact force network is generated
in response to an applied load.
We consider two-dimensional systems of frictionless discs~$\alpha$,
with radii $R_{\alpha}$ that are uniformly distributed over the interval
$[1,\Poly]$.
The parameter $\Poly>1$ will be used as a measure of the
polydispersity of the system.
The pile is built upon a fixed rough base,
consisting of a row of discs belonging to
the same size distribution as those deposited,
and whose centres all have the same vertical coordinate.
Periodic boundary conditions are obeyed in the horizontal direction.
For computational efficiency,
the beads are assumed to be weightless
except for those at the free surface, which are subjected to
a uniform load $\Load$.
Here, $\Load$ can point in a direction other than vertically downwards,
thus allowing the system to be {\em sheared}.

The pack of discs is built by adding each successive disc at the lowest
available point.  Once added, each disc's centre remains fixed throughout
the subsequent relaxation process. 
The sequential nature of the initial packing means that
it is straightforward to calculate the contact forces
as the applied load propagates from the upper surface
to the base; see \cite{adapt_net} for details on how this is done.
In general, some of the contact forces will be
tensile and thus incompatible with a cohesionless medium.
Therefore we now modify the contact network until
all tensile bonds have been removed.
This relaxation process exploits the fact
that only a fraction of the physical neighbours of a given disc are in
contact with it.  Stability is achieved by successively removing contacts
with some neighbours and adding them to others.

In a pack of discs such contact re-arrangements require motions of the discs.
To avoid such motion we make a restrictive geometric {\it Ansatz}.
We suppose that the beads once assembled have been deformed so that all
non-\-con\-tact\-ing neighbours are nearly in contact.
Specifically, we add material to each pair of non-contacting neighbours so
that the gap between them is infinitesimal.
The remaining gap is a) at the centre of the original gap, b) oriented
the same as the original gap and c) proportional to the width of the
original gap.  Under this {\it Ansatz} contact rearrangements can occur
with arbitrarily little motion.  Thus the positions of the beads and the
contact angles can be taken as fixed during the relaxation process.
The {\it Ansatz} thus simplifies the relaxation behaviour greatly.
It also abstracts the essential process of contact rearrangement from the
inessential process of bead motion.

The adaptive network approach has a number of advantages over
traditional simulation methods.
It operates directly in the limit of infinitely rigid beads,
and also in the isostatic limit~\cite{moukarzel}.
That is, the number of contacts is just sufficient to determine the
positions of the beads and the magnitudes of the contact forces. 
In two dimensions an isostatic pack has on average 4 contacts per bead.
A further advantage of using this method is that
the linear response regime can be probed directly,
a feature that will be exploited in Sec.~\ref{s:response}
of the current work.
However, the computational resources required do not scale favourably with
the number of beads~$N$, with the memory needed varying as $O(N^{2})$
and the simulation time increasing as $O(N^{4})$.
Thus its benefits are restricted to somewhat small piles,
typically $N\leq500$ here.

\begin{figure}
\centerline{\psfig{file=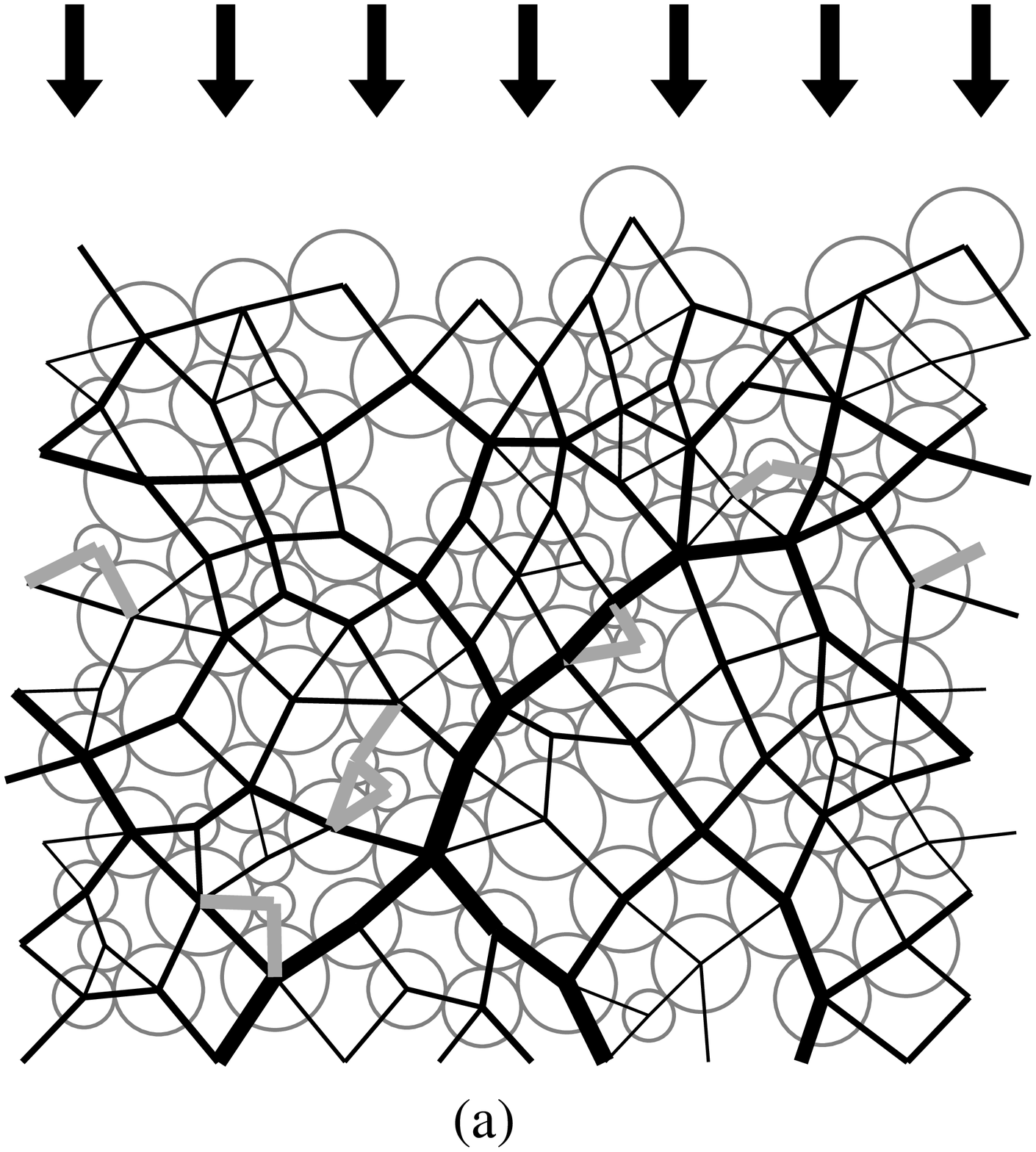,width=3in}}
\centerline{\psfig{file=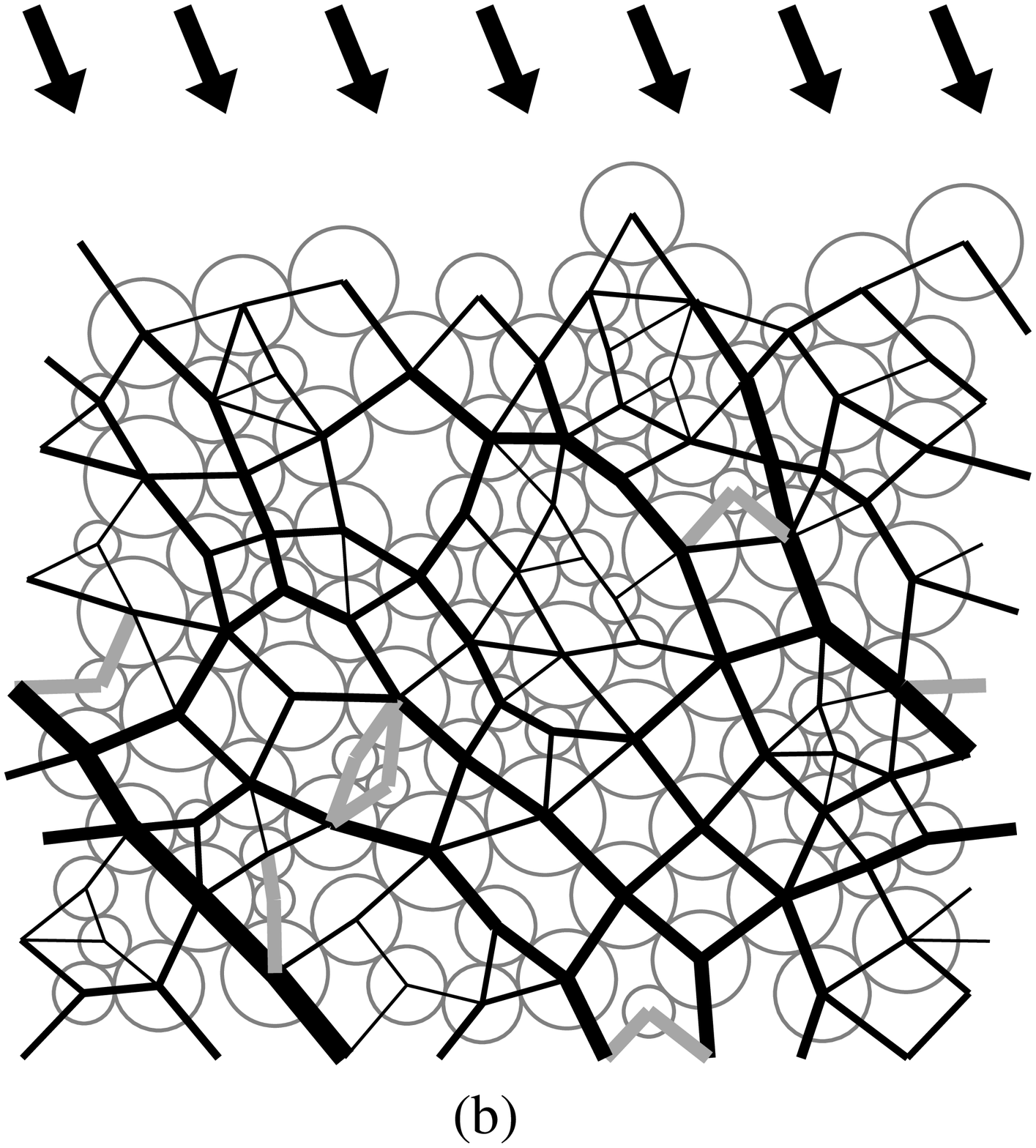,width=3in}}
\caption{Sample output of the adaptive network algorithm for a system
subjected to (a) a vertical load, and (b) for the same packing
under a shear of approximately 20$^{o}$.
The arrows point in the direction of the loading force $\Load$\,.
The black lines between beads denote compressive contacts,
with a force whose magnitude is proportional to the thickness of the line.
The thick gray lines represent tenuous contacts of exactly zero force.
Note that contacts between slightly separated beads are allowed
within the approximations of the adaptive network algorithm.
This system is small, consisting of only $N=100$ beads,
but has a high polydispersity of $\Poly=3$.
See also~\cite{URL}.
}
\label{f:examples}
\end{figure}

A sample output from our simulations is given in Fig.~\ref{f:examples},
which shows a highly polydisperse system with $\Poly=3$
that is first stabilised under a vertical load,
and then restabilised under a shearing force angled at
$20^{\circ}$ to the vertical~\cite{URL}.
The positions of the beads is preserved under the change of load;
this is the hallmark of the adaptive network algorithm.
However, the network of contact forces adjusts to support the new load,
and moreover the strongest contacts have a clear tendency to align
with the direction of $\Load$\,.
No such alignment is apparent for the weaker bonds.
The appearance of concentrated chains of force that
align in such a way as to support the load has also been observed in
experiments and simulations~\cite{bimodal,science,2dcouette}.

Before presenting our quantitative results,
we remark that although the contact forces must be non-tensile,
they need not be compressive.
Highlighted in Fig.~\ref{f:examples} are a small number
of contacts with {\em exactly} zero force.
It should be recalled that the beads in the bulk are weightless.
In a real packing, these zero-force bonds would be replaced by weak
compressive contacts that are proportional to the weight of a single bead.
Physically, these beads occupy a position in the force network that is
shielded from the applied load by an `arch' above the bead,
and thus play no role in the macroscopic propagation of the stress
through the system.
In the language of~\cite{fragility}, they are {\em spectator} beads.
The proportion of such beads is typically small but finite, roughly 2\%
for a low polydispersity $\Poly=1.1$, increasing to around 5\%
for $\Poly=3$.
There is also a slight increase in their numbers for sheared systems.

\section{The null--stress law}
\label{s:nullstress}

As mentioned in the introduction, there has been
considerable recent interest in deriving an equation
relating the components of the stress tensor for a
granular body~\cite{royalsoc,OSL,FPA,first_pre,onuttom,incipient}.
In this section we test a proposed class of equations against our simulations.
Similar results have already been published in~\cite{adapt_net},
but only for low polydispersity.
Here we extend the analysis to highly polydisperse systems
with $\Poly=3$, thus eliminating concerns over the possible
effects of crystallinity.

The stress tensor $\sigma_{ij}$ generally has 3 independent components
in two dimensions
(the absence of local torque fixes $\sigma_{xy}=\sigma_{yx}$).
Since the forces must balance at every point in the system,
$\sigma_{ij}$ obeys the stress continuity equations

\begin{equation}
\partial_{i}\sigma_{ij}=F_{j}^{\rm ext}\quad.
\label{e:stresscty}
\end{equation}

\noindent{}This gives two equations in three unknowns, so an extra
equation is required for closure.
One proposal for this `missing' equation,
called the `oriented stress linearity' model~\cite{royalsoc}
or `null stress' law~\cite{first_pre},
is that there exists a linear relationship between the
$\sigma_{ij}$,

\begin{equation}
\frac{\sigma_{xx}}{\sigma_{yy}} =
\eta + \mu \frac{\sigma_{xy}}{\sigma_{yy}}\quad.
\label{e:ceqn}
\end{equation}

\noindent{}The coefficients $\eta$ and $\mu$ depend on the construction
history of the pile, so $\mu$ will vanish if the construction respects
left-right symmetry, for example.

\begin{figure}
\centerline{\psfig{file=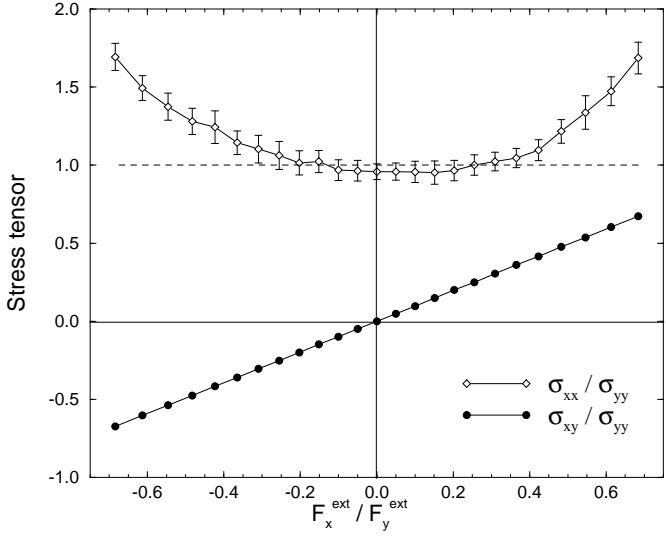,width=4in,angle=270}}
\caption{The components of the stress tensor $\sigma_{ij}$
as a function of the applied shear
for an $N=400$ bead system of polydispersity $\Poly=3$.
The error bars for the $\sigma_{xx}/\sigma_{yy}$ points
are the standard deviation over 25 different packings.
For comparison, the dashed line represents $\sigma_{xx}=\sigma_{yy}$.
The $\sigma_{xy}/\sigma_{yy}$ data lie on a line of slope 1
and the errors are smaller than the symbols.
}
\label{f:null_stress}
\end{figure}

To test the proposed equation~(\ref{e:ceqn}), we
plot in Fig.~\ref{f:null_stress} the components of $\sigma_{ij}$
for systems under differing degrees of shear.
Here the $\sigma_{ij}$ have been averaged over the bulk
of the system.
This is a valid procedure because,
with our geometry and boundary conditions,
the $\sigma_{ij}$ are constant everywhere except the narrow surface layer.
It is immediately apparent from the data that
$\sigma_{xy}/\sigma_{yy}=F_{x}^{\rm ext}/F_{y}^{\rm ext}$
to very good precision.
This is a trivial result that is easily derived from~(\ref{e:stresscty}),
and can be regarded as a consistency check for our simulations.

More revealing is the data for normal components of the stress,
which approximately obey $\sigma_{xx}=\sigma_{yy}$
for unsheared or slightly sheared systems.
This suggests that $\mu=0$, confirming that our construction
procedure preserves left--right symmetry.
The linear relationship between $\sigma_{xx}$ and $\sigma_{yy}$
breaks down under higher angles of the applied load,
indicating that the shear has changed the structure
of the system in such a way that $\mu$ is no longer zero,
and left-right symmetry is violated.
(Note that the response within a system that has been stabilised
under a large angle load is still linear, as indeed it must
always be for any packing constructed
by the adaptive network algorithm).
Equation~(\ref{e:ceqn})
may hold even for large angle shears
if $\mu$ is some function of the
ratio $F_{y}^{\rm ext}/F_{x}^{\rm ext}$,
but our data is too noisy to suggest a meaningful fit.

Additionally, the finding that $\eta\approx1$ is in agreement with
earlier work on weakly polydisperse systems~\cite{adapt_net}.
This suggests that the `speed of light' for stress propagation
is approximately~1, irrespective of polydispersity.
The analysis of the response function presented in the next section
also agrees with this observation.
We postpone further discussion of this point until Sec.~\ref{s:disc}.

\section{Response function}
\label{s:response}

The previous section considered only the coarse-grained
components of $\sigma_{ij}$ averaged over the whole bulk of the packing.
To gain further insight into the nature of the stress propagation,
we now turn to consider the response of the system to an infinitesimal
force ${\bf F}_{\gamma}$ applied to a single bead $\gamma$.

The prediction of the equation~(\ref{e:ceqn}) for
this problem is already well documented
(see {\em e.g.}~\cite{OSL,first_pre}),
but in brief, the substitution of~(\ref{e:ceqn})
into the stress continuity equations~(\ref{e:stresscty}) gives

\begin{eqnarray}
\partial_{x}(\eta\sigma_{yy}+\mu\sigma_{xy})
+\partial_{y}\sigma_{xy}&=&F_{x}^{\rm ext}\quad,\\
\partial_{x}\sigma_{xy}+\partial_{y}\sigma_{yy}&=&F_{y}^{\rm ext}\quad.
\end{eqnarray}

\noindent{}After some straightforward manipulations,
these can be rewritten as

\begin{equation}
(\partial_{y}-c^{+}\partial_{x})
(\partial_{y}-c^{-}\partial_{x})
\sigma_{ij}=0\quad,
\label{e:hyperbolic}
\end{equation}

\noindent{}where $c^{\pm}=\frac{1}{2}(-\mu\pm\sqrt{\mu^{2}+4\eta})$.
These equations are hyperbolic, and as such belong to the
same class of equations as the wave equation.
Thus they predict that the response to the perturbation should
propagate downwards only along the two characteristics,
or `light rays,' defined by $\Delta y=c^{\pm}\Delta x$,
where $(\Delta x,\Delta y)$ is the relative displacement
from the point of the perturbing force.

If this picture is correct,
then only those beads $\alpha$ and $\beta$ with a point of
contact that lies near to one of the light rays can have
their contact force altered by the application of the
perturbation.
To test this prediction, we define the response function $\Resp$ by

\begin{equation}
\Delta f_{\alpha\beta} = \Resp\cdot{\bf F}_{\gamma}\quad,
\label{e:def_of_G}
\end{equation}

\noindent{}where $\Delta f_{\alpha\beta}$ is the change in the magnitude of
the contact force between beads $\alpha$ and $\beta$ due to the
perturbing force~${\bf F}_{\gamma}$ acting on bead $\gamma$\,.
Thus $\Resp$ encodes the response to the force ${\bf F}_{\gamma}$
without requiring the direction of ${\bf F}_{\gamma}$ to be specified.
Note that for our geometry--preserving procedure, 
only the magnitude of the contact force can be varied,
as the angles were fixed during the initial preparation stage.

\begin{figure}
\centerline{\psfig{file=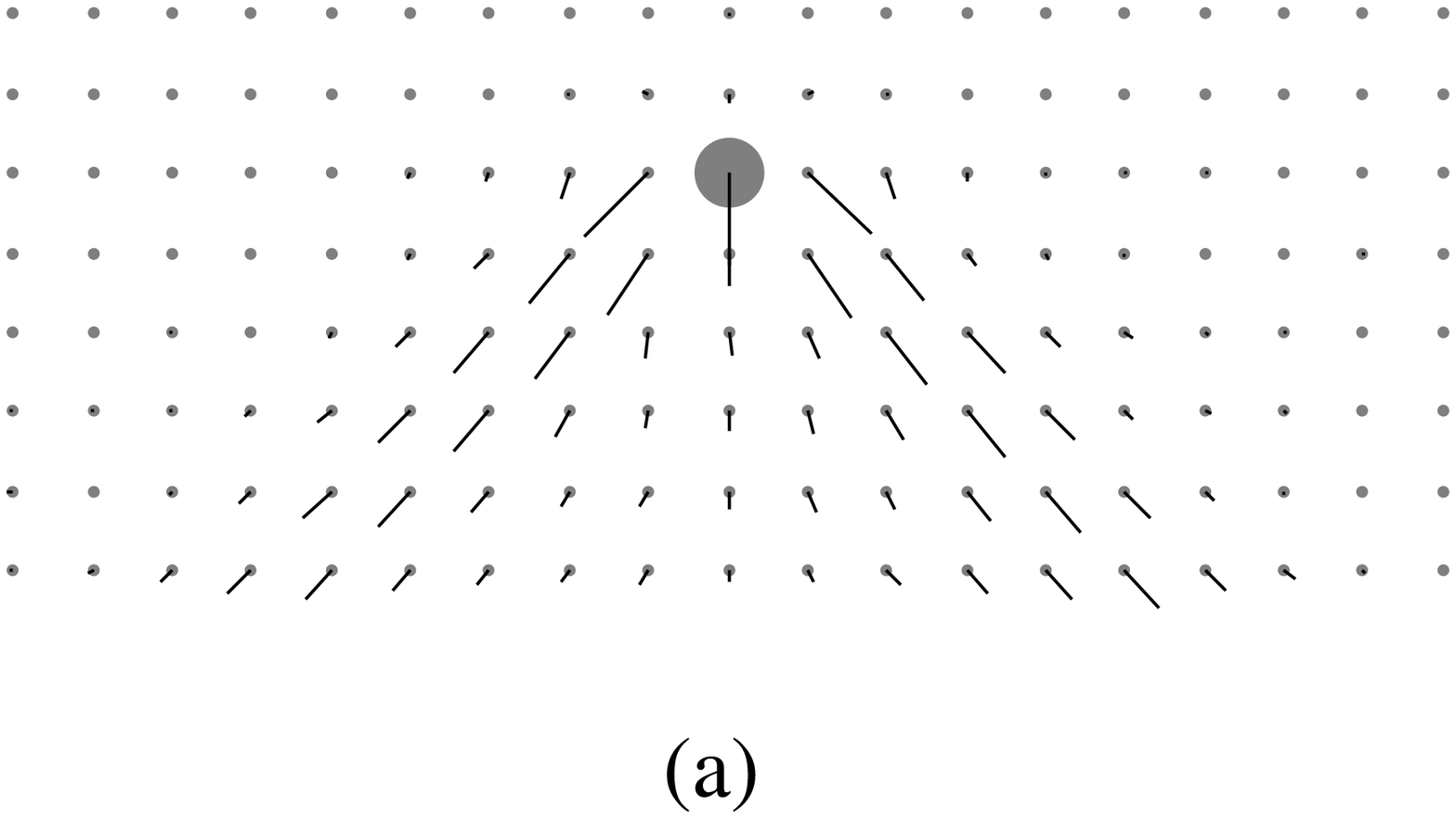,width=3.9in}}
\centerline{\psfig{file=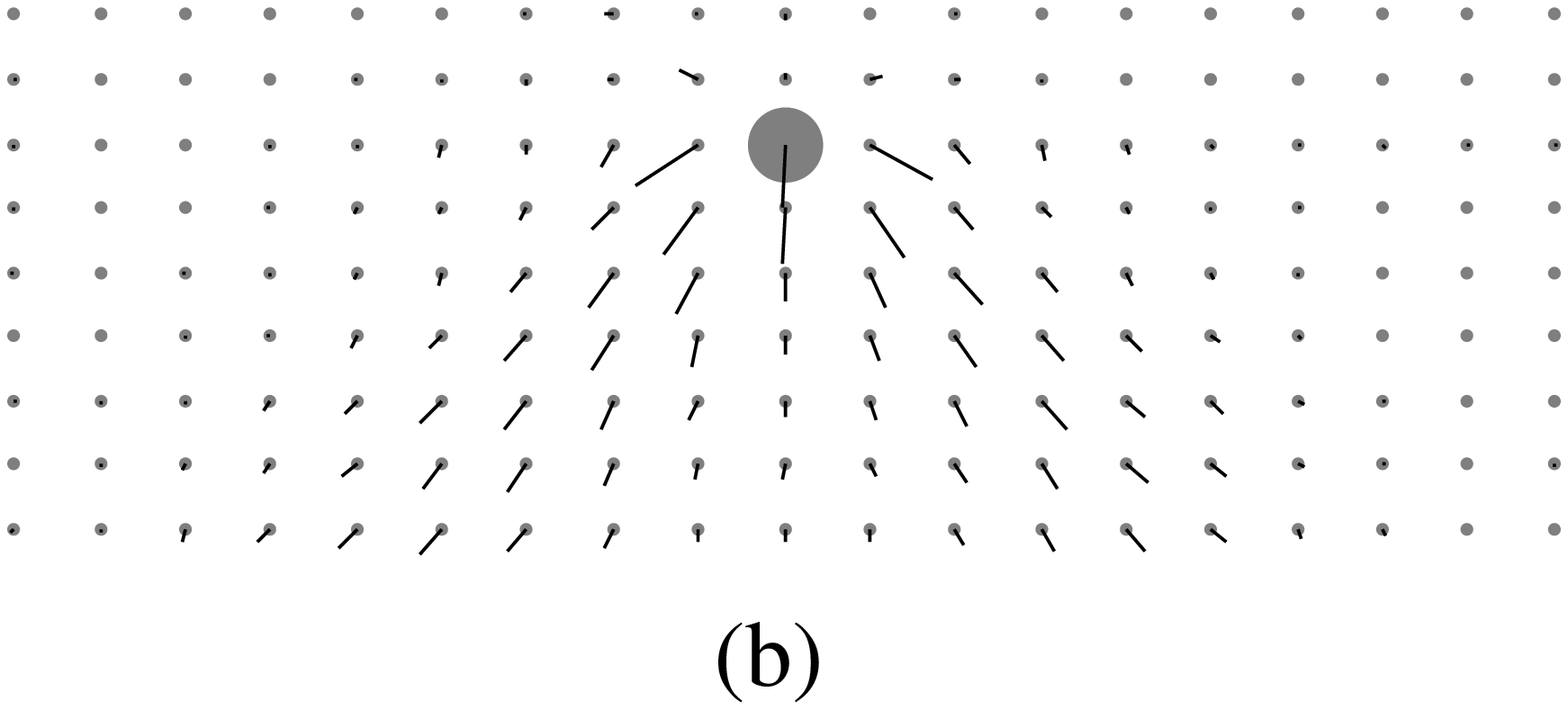,width=3.9in}}
\centerline{\psfig{file=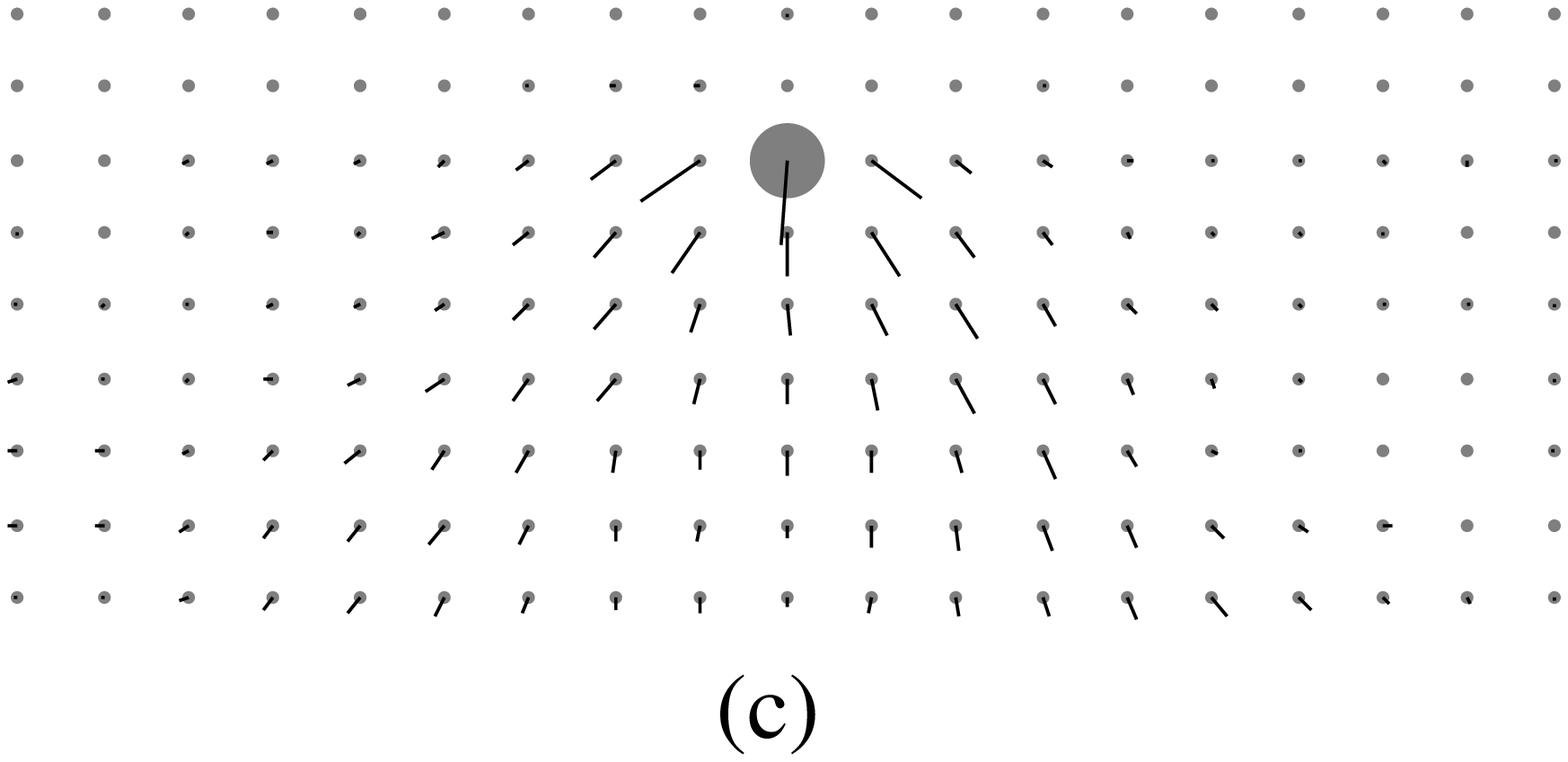,width=3.9in}}
\caption{The coarse-grained response function $\Resp$ for packings
under a vertical load and with polydispersity
(a)~$\Poly=1.1$, (b)~$\Poly=1.5$ and (c)~$\Poly=3$.
Here, $\Resp$ has been averaged over all those beads $\gamma$ whose
centres lie in a narrow horizontal strip approximately two thirds
of the way up the pile.
This central bead is represented by the large grey filled circle.
The vectors shown represent $\Resp$ as a function of the relative
displacement $(\Delta x,\Delta y)$ from the centre of bead $\gamma$
to the point of contact between beads $\alpha\leftrightarrow\beta$.
Each vector points away from the small grey circle.
The data was averaged over 100 runs of $N=500$ beads.
Each mesh point is separated by 1.5 bead diameters.
}
\label{f:resp_vert}
\end{figure}

$\Resp$ is plotted in Fig.~\ref{f:resp_vert} for systems of three
different polydispersities that have been stabilised under
a vertical load.
For the lowest polydispersity $\Poly=1.1$ the response is concentrated
along rays radiating at $\approx45^{\circ}$ angles,
suggesting that $c^{\pm}\approx \pm1$.
This cannot be due to crystallinity, as the contacts in a
close packed two-dimensional packing lie rather at a 30$^{\circ}$ angle.
For higher polydispersities, the light rays become
less distinct due to the increased noise,
and the magnitude of $\Resp$ decays more rapidly away from bead~$\gamma$.
However, on length scales of $\sim$5 beads diameters or more
it is still possible to discern
maxima in $\Resp$ along the same $45^{\circ}$ lines as before.

\begin{figure}
\centerline{\psfig{file=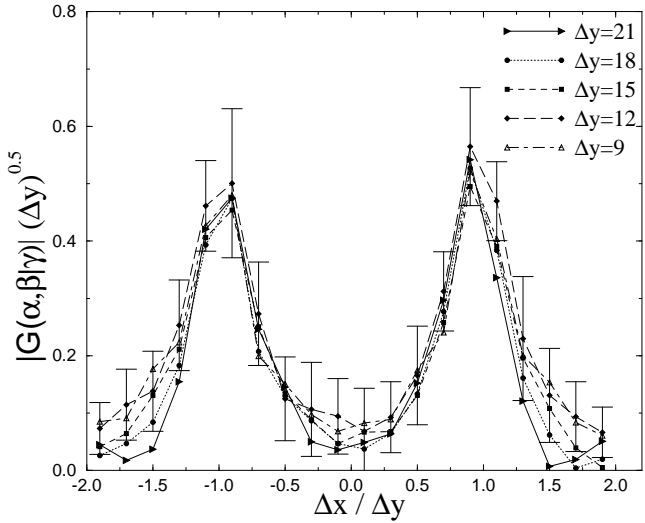,width=3.9in}}
\centerline{\psfig{file=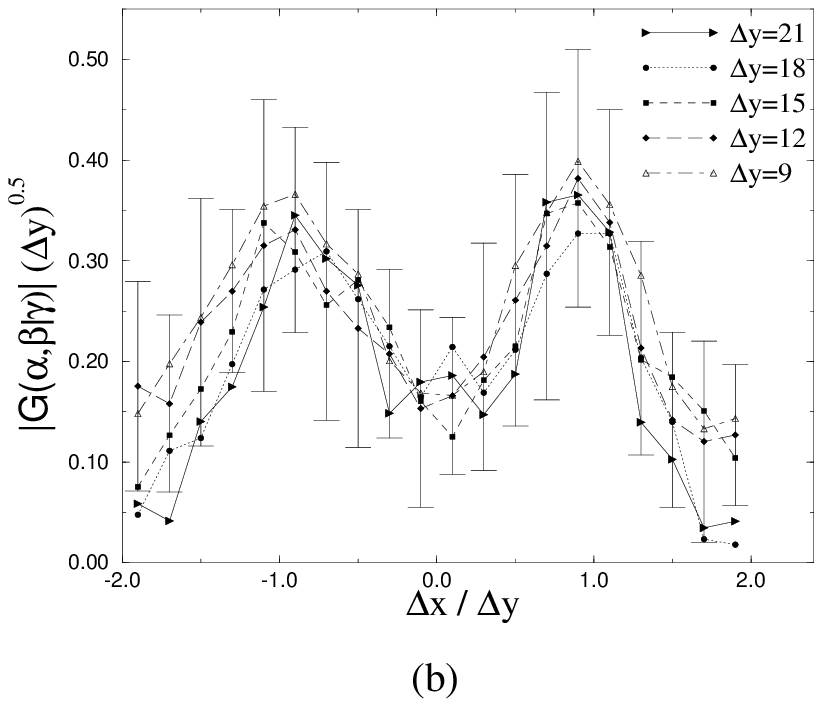,width=3.9in}}
\caption{The magnitude of the response function
$|{\bf G}(\alpha,\beta|\gamma)|$ scaled up by a
factor $(\Delta y)^{1/2}$, plotted against $\Delta x/\Delta y$,
where  $(\Delta x,\Delta y)$ is the relative displacement from
the source of the perturbation measured in units of the
mean bead radius.
Positive $\Delta y$ correspond to points {\em below} the source.
Each line represents the average over a horizontal strip of
width 3, with a mean $\Delta y$ as indicated in the legend.
The polydispersity was (a)~$\Poly=1.1$ and (b)~$\Poly=1.5$,
and $N=500$ in both cases.
For clarity, only the error bars for the $\Delta y=12$ line are shown.
These error bars give the standard deviation of 10 data points, each of
which is the mean of 10 separate runs.
The $\Poly=3$ data is not given as the error bars were much larger
than the data points.
}
\label{f:quant_vert}
\end{figure}

To make these observations more quantitative,
we plot in Fig.~\ref{f:quant_vert} the magnitude of the response
$|{\bf G}(\Delta x,\Delta y)|$
at different depths $\Delta y$ below the perturbation.
It is evident that the peak response lies near to the lines
$\Delta y=\pm\Delta x$, again confirming that $c^{\pm}\approx\pm1$.
Furthermore, the amplitude of the peaks decays like
$(\Delta y)^{-1/2}$, which
presumably arises from the diffusive spreading--out of
$|{\bf G}(\Delta x,\Delta y)|$ with increasing depth~\cite{CD}.
Diffusive spreading should also mean that the width of the peaks
increases like $(\Delta y)^{1/2}$, but our data is too noisy
to test this claim at present.

It is interesting to compare our results to the recent model
of Bouchaud {\em et al.} which allows the force to
propagate along straight rays until they are scattered by a random
distribution of defects~\cite{splaymodel}.
These authors found that disorder causes a crossover from a two--peak response
function on short length scales to a single--peak, `pseudo-elastic'
behaviour on longer scales.
By contrast, we have found no evidence of any crossover, and the
two-peak form of $|{\bf G}(\Delta x,\Delta y)|$ persists to the
longest length scales we have been able to simulate.
It is not clear if this discrepancy has anything to do with the
differing underlying assumptions between the two models,
or if we are simply unable to generate sufficiently large systems
to observe the crossover.

\begin{figure}
\centerline{\psfig{file=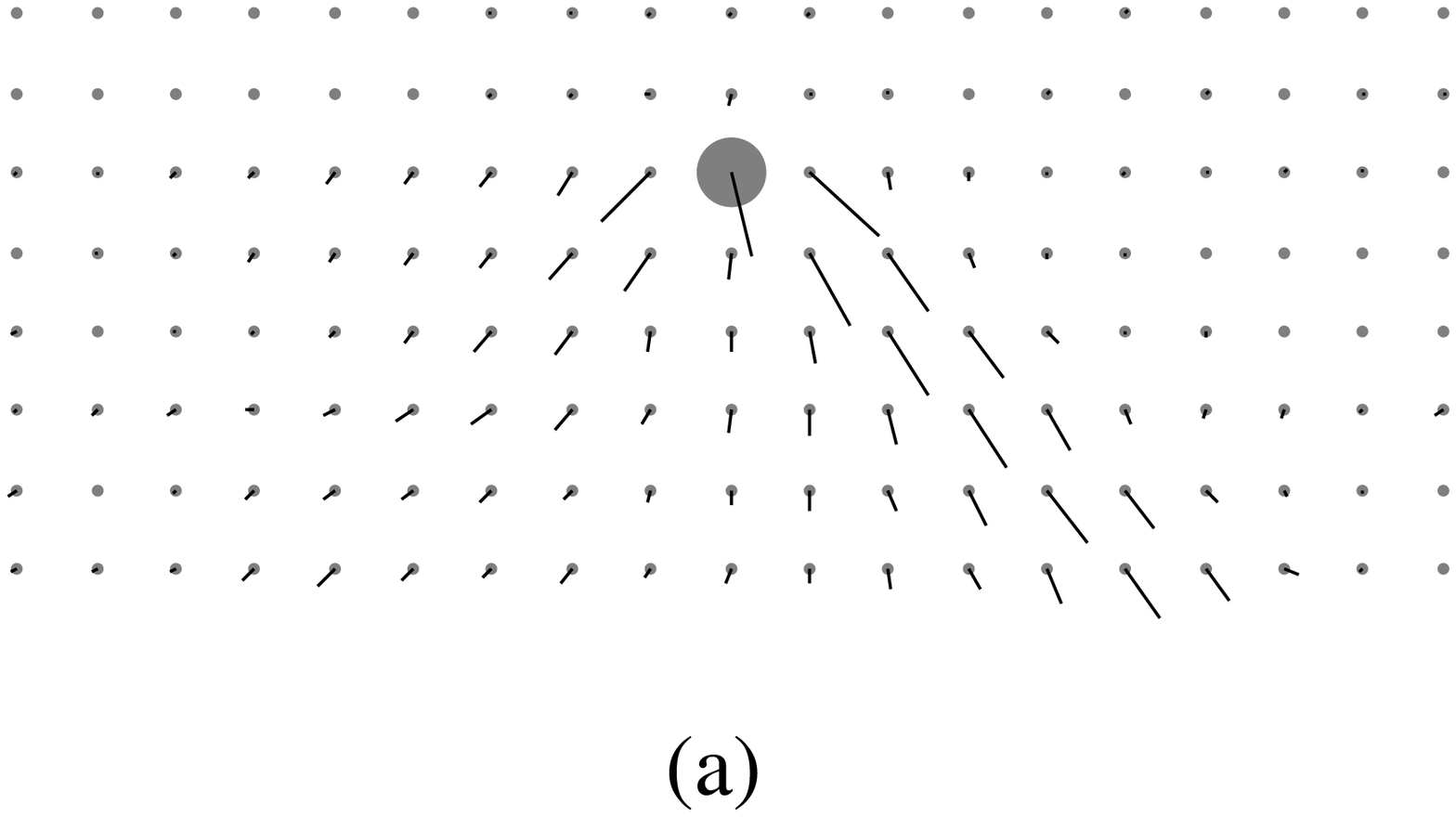,width=3.9in}}
\centerline{\psfig{file=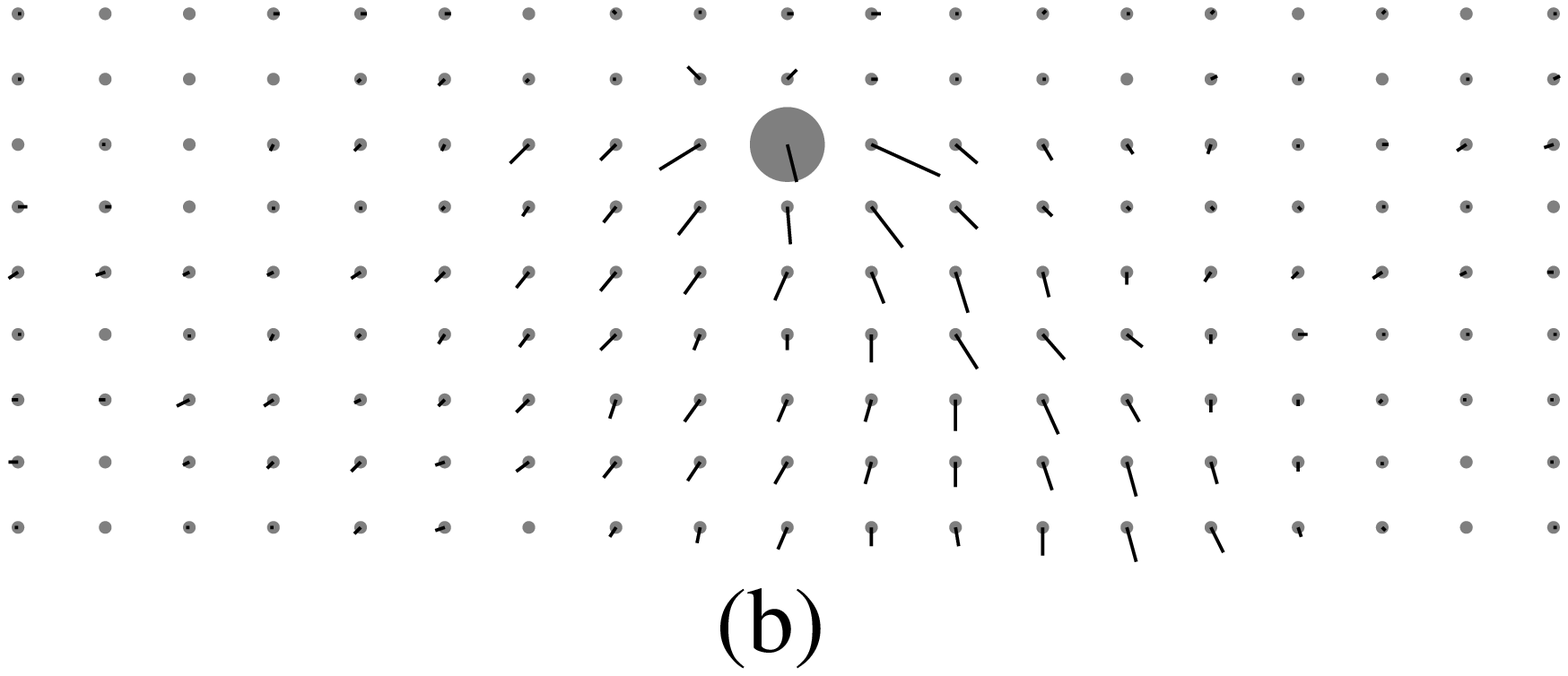,width=3.9in}}
\centerline{\psfig{file=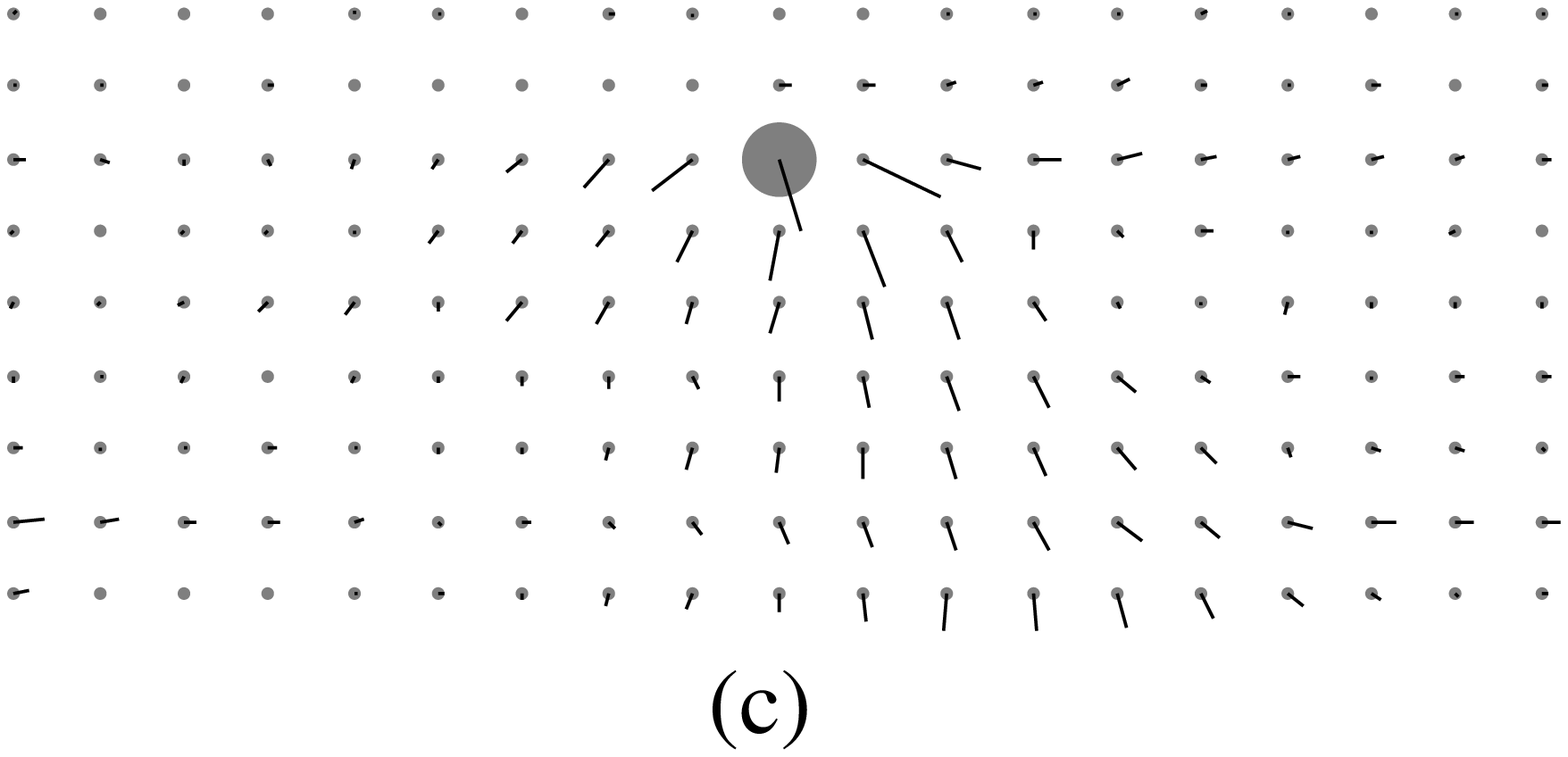,width=3.9in}}
\caption{The response function $\Resp$ when the applied load is
angled at $20^{\circ}$ to the right of the vertical,
for polydispersities
(a)~$\Poly=1.1$, (b)~$\Poly=1.5$ and (c)~$\Poly=3$.
}
\label{f:resp_shear}
\end{figure}

\begin{figure}
\centerline{\psfig{file=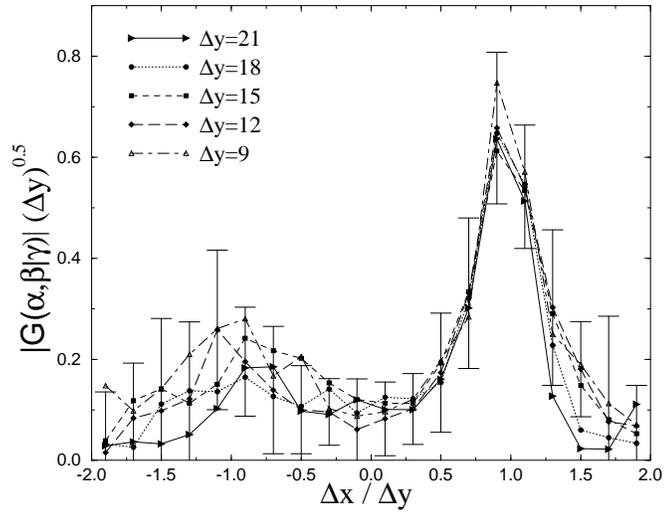,width=3.9in,angle=270}}
\caption{The same as Fig.~\ref{f:quant_vert} for a system
under a shearing load, angled at $20^{\circ}$ to the vertical.
Only the results for weakly polydisperse system $\Poly=1.1$ is given,
as the error bars were much larger than the data points for
higher polydispersities.
}
\label{f:quant_ang}
\end{figure}

The response function for systems that have been stabilised
under a shearing load are given in Figs.~\ref{f:resp_shear}
and~\ref{f:quant_ang}.
It is readily apparent that applying the shear does not
alter the orientation of the light cone,
{\em i.e.}~the characteristics have the same gradients $c^{\pm}\approx\pm1$
as in the unsheared system.
However, the magnitude of the response becomes stronger along
the ray that extends in the same direction as the shear,
and weaker along the opposite-pointing ray,
as required by~(\ref{e:def_of_G}).
With increased polydispersity, there is a greater degree of noise
and the weaker ray is eventually lost in the background,
as seen in {\em e.g.} Fig.~\ref{f:resp_shear}c.

\section{Discussion}
\label{s:disc}

The central finding of this paper has been the repeated observation
of a `speed of light' $c=1$ for the characteristics along
which the stress propagates.
This was first inferred from the form of the `constitutive equation'
described in Sec.~\ref{s:nullstress}, which for unsheared or
mildly sheared systems were consistent with
$\sigma_{xx}=\eta\sigma_{yy}$ with $\eta=c^{2}\approx1$.
Confirmation came from the response functions in Sec.~\ref{s:response},
which were concentrated
within a `light cone' with a central axis running parallel to gravity.
The gradient of the surfaces of this cone were again $c^{\pm}\approx\pm1$.
In other words, the  two characteristics of the hyperbolic equation for
stress are nearly orthogonal. 
Our earlier study \cite{adapt_net} indicates that this property
survives even when the preparation procedure is not mirror--symmetric. 
In that case, the null--stress law has the general form
$\sigma_{xx}=\eta \sigma_{yy} +\mu \sigma_{xy}$,
with the value of $\eta$ still consistent with~$1$.
As seem from~(\ref{e:hyperbolic}),
if $\eta=1$ then the characteristics obey
$c^{+}c^{-}=-1$ for all $\mu$, {\em i.e.} they are orthogonal,
just as in the fixed principle axes theorem~\cite{FPA}.
The fact that the orthogonality of the
characteristics appears to be so robust, demands an explanation.

Equality between the normal stresses $\sigma_{xx}=\sigma_{yy}$
is also found in liquids, where it reflects the isotropic
nature of the medium.
However, this interpretation cannot be applied to our model,
since the construction history and boundary conditions give rise
to a system that is anything but isotropic.
The anisotropy is most explicit in the response functions
plotted in {\em e.g.} Fig.~\ref{f:resp_vert}, which clearly
show that the orientation of the light cone is fixed with respect
to gravity, irrespective
of the direction of the perturbing force~${\bf F}_{\gamma}$
or the degree to which the system is being sheared.
Clearly we must look elsewhere for an explanation.

In the absence of a more compelling reason why $c=1$, we tentatively
propose the following explanation.
Firstly, observe that the relation $\sigma_{xx}=\sigma_{yy}$
means that the volume of the system is conserved under small perturbations.
To see this, consider a system that is contained
within a rectangular box of width $X$ and height $Y$.
Let us suppose that, under the action of an external force
$f_{x}$ applied to one of the side walls, the width of the system
changes to $X+\delta X$.
Similarly, a force $f_{y}$ is applied to the top of the box,
and the height changes to $Y+\delta Y$.
If both $\delta X$ and $\delta Y$ are sufficiently small,
the topology of the contact network is not altered during this
motion and the perturbation is a `soft mode'~\cite{first_pre},
meaning that it does not change the total energy of
the system and hence has no restoring force.
This follows from the isostatic nature of the network;
see~\cite{first_pre} for details.
To avoid runaway motion, the total work done on the system
must vanish, {\em i.e.} $f_{x}\,\delta X+f_{y}\,\delta Y=0$.
But the relation $\sigma_{xx}=\sigma_{yy}$ can be rewritten
as $f_{x}/Y=f_{y}/X$, giving $Y\delta X+X\delta Y=0$.
The left hand side of this equation is just the change in volume of the system,
which therefore remains fixed, as claimed.

It is plausible to suppose that our observation of $c=1$ arises from a state 
of minimal volume or maximal compaction, as suggested above.  This suggests 
that $c$ may be unity in a broader class of physical systems.  Still, we 
have not demonstrated the necessary conditions for a minimal-volume packing.
It may require a random close--packed initial state.  It also may require 
frictionless contacts or regular bead shapes.  Further, we have not 
demonstrated that our simulated bead pack has minimal volume in the 
requisite sense.  Nor have we shown that it would continue to have $c=1$ if 
we allowed the beads to move during relaxation.
Thus the class of systems with $c=1$ beyond that presented here
cannot be fully characterised at this time.

\section{Summary}
\label{s:conc}

In summary, we have investigated the propagation of stress
through a body of frictionless rigid discs using the adaptive
network algorithm.
This algorithm first generates a two--dimensional
packing via a sequential deposition procedure,
and then calculates the contact forces in response
to a load applied to the free surface.
Tensile bonds are removed by altering the topology of
the contact network, in a manner that preserves
the mean coordination number;
the packing is always isostatic.
This is performed in the linear regime in which bead deformation
and motion is neglected, so that although
the topology of the network can evolve,
the geometry is preserved.

The linear response to a force applied to a single bead
was calculated, and was found to propagate in two
downward--pointing `rays,'
Our central result is that the `speed of light' $c$ of
these rays is approximately~1,
irrespective of the polydispersity of the
bead size distribution,
or the degree to which the system is being sheared.
We have also argued that $c=1$ may be understandable
if the system naturally adopts a local minimum of volume.
Furthermore, we have shown that the magnitude of the
response function decays in a way that suggests
a diffusive spreading--out with the vertical distance
from the source of the perturbation.
It would be interesting to see if other microscopic models
also scale in a similar manner.

There are many possible directions in which the results
presented in this paper can be extended.
In particular, we feel that there would be considerable advantages
to devising a modified adaptive network algorithm that scales more
favourably with system size, whilst still directly probing the
linear response regime.
This would help to reduce the significant error bars found
in some of our data.
One way to achieve this might be to change the rule that updates
the system when a single bond is changed,
which is currently global and requires $O(N^{2})$ calculations.
Replacing this with a local rule should significantly improve
the simulation times and hence the statistics,
although it is not clear what approximations such an optimisation would entail.

The current work has focused purely on the problem of stress propagation
through granular materials.
However, granular media make up only a small subclass of systems
that can be described as `jammed,'
a class that also includes glasses and foams~\cite{corey}, for example.
It has recently been demonstrated that the distribution of
interbead forces in these systems exhibits a kind
of universality when in their jammed state~\cite{corey}.
Clearly the adaptive network algorithm could be a useful probe
of the extent and limits of any such universality.
Since this is a separate question to that considered in this paper,
we will present our results on this issue elsewhere~\cite{nextpaper}.

\section*{Acknowledgements}
\label{s:ack}

The authors would like to thank Mike Cates and Joachim Wittmer
for informative discussions regarding the work presented in this paper.
DAH was funded by EPSRC (UK) grant no. GR/M09674.
This research was supported in part by the US National Science Foundation
through its MRSEC program under award DMR-980859.



\end{document}